\documentstyle[aps,epsf,graphicx]{revtex}
\def\xvec{{\bf x}}

\def\uvec{{\bf u}}
\def\vvec{{\bf v}}
\def\nuvec{{\mbox{\boldmath $\nu$}}}

\def\Omvec{{\mbox{\boldmath $\Omega$}}}
\def\crossprod{{\mbox{\boldmath $\times$}}}

\def\xivec{{\mbox{\boldmath $\xi$}}}
\def\grad{{\mbox{\boldmath $\nabla$}}}

\def\be{\begin{equation}}
\def\ee{\end{equation}}
\def\baray{\begin{eqnarray}}
\def\earay{\end{eqnarray}}
\def\bnabla{{\mbox{\boldmath $\nabla$}}}
\newcommand{\bDelta}{{{\bf \Delta}}}

\begin{document}

\title{Oscillations of rotating trapped Bose-Einstein condensates}

\author{A. Sedrakian$^{1)}$ and I. Wasserman$^{2)}$}
\address{
$^{1)}$ Kernfysisch Versneller Instituut,
                             NL-9747 AA Groningen, The Netherlands \\
$^{2)}$ Center for Radiophysics and Space Research,
                             Cornell University, Ithaca, NY 14853
          }

\maketitle


\begin{abstract}
The tensor-virial method is applied for a study of oscillation
modes of uniformly rotating Bose-Einstein condensed gases, whose
rigid body rotation is supported by an vortex array.
The second order virial equations are derived in
the hydrodynamic regime for an arbitrary external harmonic
trapping potential assuming that the condensate is a
superfluid at zero temperature. The axisymmetric equilibrium
shape of the condensate is determined as a function of the
deformation of the trap; its domain of stability is bounded
by the constraint $\Omega<1$ on the rotation rate
(measured in units of the trap frequency $\omega_0$.)
The oscillations of the axisymmetric condensate are stable
with respect to the transverse-shear and toroidal
modes of oscillations, corresponding to the $l= 2$,
$\vert m\vert = 1,2 $ surface deformations.
The eigenfrequencies of the modes are real and represent
undamped oscillations.
The condensate is also stable against quasi-radial pulsation
modes ($l=2$, $m=0$), and its oscillations are undamped,
if the superflow is assumed incompressible. In the
compressible case we find that for a  polytropic
equation of state, the quasi-radial oscillations are unstable
when $\gamma(3-\Omega^2)<1-3\Omega^2$, and are stable
otherwise. Thus, a dilute Bose gas, whose equation of
state is polytropic with $\gamma=2$ to leading order in
the diluteness parameter, is stable irrespective of the
rotation rate. In non-axisymmetric traps, the equilibrium
constrains the (dimensionless) deformation in the plane orthogonal
to the rotation to the domain $A_2 > \Omega^2$ with
$\Omega< 1$. The second
harmonic oscillation modes in non-axisymmetric traps separate
into two classes which have even or odd parity with respect
to the direction of the rotation axis. Numerical solutions
show that these modes are stable in the parameter domain where
equilibrium figures exist.
\end{abstract}

\section{Introduction}
After the experimental realization of Bose-Einstein
condensation in vapors of alkali atoms \cite{ANDERSON,BRADLEY,DAVIS},
the understanding of their behavior under rotation
became the main focus of theoretical and experimental work.
Experimental observation of the vortex states in a
two-component system\cite{JILA}, and the realization of
the vortex states in a stirred one-component Bose-Einstein
condensate \cite{MADISON} confirmed the expectations
that sufficiently large condensates exhibit
bulk superfluid properties by supporting their rotation
above the critical angular frequency  by the
Feynman-Onsager vortex state\cite{BP,DALFOVO96,FETTER,FCS,SF}.

The theoretical approaches to the Bose-Einstein condensed
gases commonly distinguish between the two regimes of the
strong and the weak interparticle interactions, which
correspond to the conditions $Na/d \gg 1 $ and $Na/d \ll 1$,
respectively, where $N$ is the number of particles in the gas,
$a$ is the scattering length, and  $d = \sqrt{\hbar/m\omega_0}$
is the oscillator length defined in terms of the oscillator
frequency $\omega_0$ and the boson mass $m$ (for a review
see e.g. \cite{FETTER,DALFOVO}.)
In the weak coupling regime,
where the scale of the variation of the condensate
wave-function is of the same order as the size of the cloud,
quantum effects play an essential role and the system behaves much
like an atomic nucleus (see e. g. Ref. \cite{MOTTELSON}).
In the strong coupling regime and for sufficiently large
condensate, the condensate is well
described in terms of hydrodynamics.
Under these conditions the
time-dependent Gross-Pitaevskii (G-P) equation\cite{GP} for
the (complex) condensate wave  function reduces
to two (real) hydrodynamic
equations for the density and velocity of the condensate
at zero temperature. One consequence of the assumption that
the system is sufficiently large is that the quantum pressure term can be
ignored compared with the interaction terms. Another is that
the Thomas-Fermi approximation\cite{BP} is valid and the equilibrium kinetic
energy density is negligible compared to the potential energy density.
This approximation also ensures that the thermodynamic
quantities like the pressure or the chemical potential of the gas
are well defined at any local point of the gas cloud. Finally,
the coherence length of the condensate $\xi = 1/\sqrt{8\pi
na}$, with $n$ being the number density of condensate particles,
becomes sufficiently small in this limit to allow for a description of a
vortex core as a singularity.

The superfluid hydrodynamic approach
has been used in the  studies of the
collective excitations of the condensate clouds\cite{FETTER96,STRINGARI}
the scissors mode oscillations\cite{GOS},
the moment of inertia of a rotating condensate\cite{ZAMBELLI},
mainly in the irrotational regime; see however Ref. \cite{GRPG}.
The critical angular velocity of condensate, $\Omega_{c1}$,
at which the creation of a vortex is energetically favorable, scales,
in units of the trap frequency, as $\Omega_{c1} \propto (d^2/R^2)\, {\rm ln}
\left(R/\xi\right)$, where $R$ is the size of the cloud;
the numerical coefficient depends on
the details of the geometry of the trap\cite{FETTER}.
For a sufficiently large condensate the ground
state would correspond to coarse-grained rigid-body rotation
supported by a vortex lattice. In this case, the irrotational
constraint on the average superfluid velocity $\grad \crossprod {\bf v}=0$
is replaced by the rigid-body rotation
condition $\grad \crossprod {\bf v}=2\Omvec$.
The purpose of this paper is the study of the oscillations of the
trapped Bose condensates under rigid-body rotation. We shall not
discuss the modes related to the vortex structure
itself (e.g. Tkachenko modes).
We shall confine ourselves to the zero-temperature limit, in which case
the effects of mutual friction  due to the interaction
of the thermal excitations with the vortex state, and
dissipation due to the  viscosity of the thermal component
both can be neglected.

Below, we shall apply the tensor virial method originally
developed for the study of equilibrium
and stability of rotating liquid masses
bound by self-gravitation\cite{CHANDRA}.
The tensor virial method transforms the local hydrodynamical equations
into global virial equations that contain the full
information on the structure and stability of a system as a whole.
The method  is especially useful for studying perturbations
of incompressible uniform ellipsoids from equilibrium, in which
case the each perturbed virial equation yields (in the
absence of viscous dissipation) a different set of normal modes.
The extension to compressible flows is straightforward in the
case of adiabatic perturbations. Moreover, when the equilibrium
is sustained by an external
confining potential, fluid perturbations do not back-react (change)
the confining potential itself, and the tensor virial method can be extended to
non-uniform compressible flows. The description is particularly
useful for gases with polytropic equations of state
$p\propto \rho^{\gamma}$,
where $p$ is the pressure, $\rho$ is the
density, and $\gamma$ is the adiabatic index. In that case, we
shall see that all of the modes of the gas can be found from
the tensor virial equations. (We illustrate this only for the
second harmonic modes of oscillation in this paper, but the
extension to higher order modes is straightforward,
and will be discussed elsewhere.)
For an interacting Bose gas the pressure is a non-analytic function
of the diluteness parameter $na^3$. However the equation of
state to the leading order in the small parameter  $na^3\ll 1$ can
be written in a polytropic form: $p= K \rho^{2}$, with $K =
2\pi\hbar^2 a/m^2$. Since the zero temperature
equations of motion of a trapped rotating
condensate turn out formally identical to the
corresponding equations of motion of a ordinary (non-superfluid) liquid,
our results might be of significance in a broader context. On the other
hand, present results can serve as a starting point for an extension to
finite temperatures where, in addition to
the superfluid, the fluid of normal quasiparticle excitations plays a role.

The paper is organized as follows. In \S~2 we derive the first and
second order virial equations for trapped Bose-condensed gases
in the hydrodynamic regime. The equilibrium shape of rotating
condensates  and the  second harmonic modes of
oscillations in axisymmetric traps are discussed in \S~3.
\S~ 4 discussed the second harmonic modes of oscillations in
non-axisymmetric traps. Our results are summarized in \S~5.
The Appendix gives a brief derivation of the hydrodynamic equations
from the G-P equation.

\section{Virial equations for a trapped condensate }
\label{ve2pe}

The Euler equation for a condensate in a harmonic, arbitrary
deformed trap, can be written as
\be
\rho \left({\partial\over\partial t}+u_{j} {\partial\over\partial x_j}
\right)  u_{i}=-{\partial p\over
\partial x_i}-\frac{\rho}{2}
{\partial\phi_{\rm tr}\over\partial x_i}
+\frac{\rho}{2}{\partial\vert\Omvec\crossprod\xvec\vert^2
\over\partial x_i}+2\rho\epsilon_{ilm}u_{l}\Omega_m,
\label{eq:euler}
\ee
where the Latin subscripts denote coordinate directions; $\rho$,
$p$, and ${\bf u}$ are the density, pressure, and
velocity of condensate; (summation over repeated indexes is assumed).
The external harmonic trapping potential is
\be
\phi_{\rm tr} (\xvec) = \omega_0^2\sum_{i=1}^3 A_i x_i^2,
\ee
where $A_i$ are the dimensionless deformation parameters,
 $\omega_0$ is the frequency of the
harmonic oscillator in the trap.
Eq. (\ref{eq:euler}) is written in a frame rotating with
angular velocity $\Omvec$ relative to some inertial coordinate
reference system. Its derivation from the G-P equation
for the condensate wave-function is given in the Appendix. Our
fundamental assumption is that the condensate undergoes a rigid-body
rotation, which is supported by an array of vortices. In this case, as explained
in the Appendix, the equations of motion are formally identical to
those of a ordinary compressible (non-superfluid) liquid.

Starting from Eq. (\ref{eq:euler}) we can construct a
set of virial equations of various order.
Suppose that the condensate occupies a volume $V$.
Taking the zeroth moment of Eq. (\ref{eq:euler})
amounts to integrating over $V$; doing so, we obtain
\baray
{d\over dt}\biggl(\int_{V}{d^3x\rho u_{i}}
\biggr)&=&2\epsilon_{ilm}\Omega_m\int_{V}{d^3x\rho
u_{l}}+(\Omega^2\delta_{ij}-\Omega_i\Omega_j)
\int_{V}{d^3x\rho x_j}
\nonumber\\& &
-\int_{V}{d^3x\frac{\rho}{2}{\partial\phi_{\rm tr}\over\partial x_i}}.
\label{eq:v1}
\earay
The first order virial equation describes the
center of mass motion of the condensate, and is trivial as
these motions correspond to a uniform translation of the
system as a whole.

Taking the first moment of Eq. (\ref{eq:euler}) results in the
second order virial equation
\baray
{d\over dt}\biggl(\int_{V}{d^3x\rho x_j u_{i}}
\biggr)&=&2\epsilon_{ilm}\Omega_m\biggl(\int_{V}
{d^3x\rho x_ju_{l}}\biggr)
+(\Omega^2 - \omega_0^2 A_i) I_{ij}-\Omega_i\Omega_kI_{kj}
\nonumber\\& &
+2{\cal T}_{ij}+\delta_{ij}\Pi,
\label{eq:v2}
\earay
where
\baray
I_{ij}\equiv\int_{V}{d^3x\,\rho x_ix_j}, \quad
\Pi \equiv\int_{V}{d^3x\,p}, \quad
{\cal T}_{ij}\equiv {1\over 2}\int_{V}{d^3x\rho
u_{i}u_{j}}.
\earay

Consider the variation of the second order virial equation
under the influence of perturbations. The Eulerian variations
of various terms in Eq. (\ref{eq:v2}) are straightforward\cite{VARIATIONS}
and we find
\be
\delta \frac{d}{dt} \int_{V}d^3x \rho u_i x_j =
(\Omega^2 -\omega_0^2 A_i) V_{ij}
 - \Omega_i\Omega_k V_{kj}
+ 2 \epsilon_{ilm}\Omega_{m} \delta \int_V d^3x
\rho u_l x_j+2\delta {\cal T}_{ij}+ \delta_{ij} \delta \Pi  ,
\ee
where for a Lagrangian displacement $\xivec$
\be
V_{ij} \equiv \delta I_{ij} = \int_V d^3x \rho
(\xi_i x_j + \xi_j x_i).
\ee
The tensors $V_{ij}$ are manifestly symmetric in their
indexes, as is evident from their definition. It is useful
to define also their non-symmetric parts as
\be
V_{i;j} = \int_{V} d^3x \rho \xi_i x_j,
\ee
such that $V_{ij} = V_{i;j}+V_{j;i}$.
When there are no condensate motions in the unperturbed
state in the rotating frame ($\uvec = 0$), the second order virial
equation becomes
\baray
{d^2V_{i;j}\over dt^2}&=&2\epsilon_{ilm}
\Omega_m {dV_{l;j}\over dt}
+(\Omega^2 -\omega_0^2 A_i) V_{ij}-\Omega_i\Omega_kV_{kj}
+\delta_{ij}\delta\Pi.
\label{eq:v3}
\earay
The tensor virial Eq. (\ref{eq:v3}) can be used
for the study of small amplitude oscillations of a
trapped condensate. For time-dependent Lagrangian
displacements of the form
\be
\xivec (x_i, t) = \xivec (x_i) e^{\lambda t}
\ee
Eq. (\ref{eq:v3}) becomes
\be
\lambda^2 V_{i;j} - 2\Omega \epsilon_{ilm}
\Omega_m \lambda V_{l;j} =
(\Omega^2 -\omega_0^2 A_i) V_{ij}-\Omega_i\Omega_kV_{kj}
+\delta_{ij}\delta\Pi.
\label{eq:v4}
\ee
Eq. (\ref{eq:v4}) contains all the second harmonic
modes of the rotating condensate in a arbitrary harmonic trap.

\section{Equilibrium and oscillations in axisymmetric traps}

\subsection{Equilibrium shape}

Next, we specialize Eq. (\ref{eq:v2}) to the case of axisymmetric
traps, $A_1 = A_2 \neq A_3$ and  assume, without loss of
generality,  $A_1 = 1$. The assumption of the axisymmetry is,
in practice,  an approximation as a perfectly axisymmetric trap
will not exert a torque on the condensate. Our assumption is
that the deviations from the axisymmetry, which drive the
rotation of the condensate, are small compared with the deformation
of the trap $\vert A_1-A_2\vert /A_1 \ll 1$. In principle,
once the condensate is brought to rotation the axisymmetry can
be restored.

Consider a equilibrium state in which the condensate
rotates uniformly with the rotation vector along the $x_3$ axis
of Cartesian system of coordinates.  Then  Eq. (\ref{eq:v2}) reduces to
\be
\Omega^2 \left(I_{ij} - \delta_{i3} I_{3j} \right)
-\omega_0^2 A_i I_{ij} = -\delta_{ij} \Pi.
\label{eq:eqbr}
\ee
In axisymmetric traps with
$a_3$ denoting the  ellipsoidal semi-axis along the rotation,
the two remaining semi-axis of the ellipsoid,
$a_1$ and $a_2$, are equal.
The diagonal component of Eq. (\ref{eq:eqbr}) provides the
relation between the rotation frequency, the deformation
parameters along the $x_3$ direction (both are fixed in an experiment)
and the ratio of the semi-axis
of the resulting spheroidal figure:
\be\label{eq:figures}
\frac{a_3^2}{a_1^2} = \frac{1}{A_3}
\left(1-\frac{\Omega^2}{\omega_0^2}\right).
\ee
Depending whether $a_3\le a_1$ or $a_3 \ge a_1$ the
condensate shape is either oblate or prolate.
Note that the rotation frequency is bounded from
above, $\Omega/\omega_0 < 1$, and  for $A_3 > 1$ the equilibrium
figures are always oblate.  A striking consequence of the
Eq. (\ref{eq:figures}),  when $A_3 = 1 - \Omega^2/\omega_0^2$,
is an equilibrium figure which is a rotating sphere\cite{REMARK}.
Although we had to assume small deviations from
axisymmetry to make the condensate rotate,
this assumption is a prerequisite for producing rotation,
but not an intrinsic property of the rotating condensate.

The equilibrium density profile of the condensate, for a
polytropic equation of state, can be obtained by a direct
integration of the unperturbed limit of Eq. (\ref{eq:euler}):
\be\label{PROFILE}
\rho(\xvec)
= \rho_0 \left\{1-\frac{\gamma-1}{2K~\gamma~\rho_0^{\gamma-1}}
\left[\left(1-\Omega^2\right)x_1^2+
\left(A_2-\Omega^2\right)x_2^2
+A_3x_3^2\right]\right\}^{\frac{1}{\gamma-1}}~ \theta[\dots ],
\ee
where $\rho_0$  is the central density of the cloud, the dots
in the argument of the $\theta$-function (which insures that
the condensate density is positive) stand for the expression
in the curly brackets. Here we have retained the parameter $A_2\neq A_1$
to incorporate the case of non-axisymmetric traps which is treated
in the next section.
The effect of the rotation is the stretching of the profile of the
condensate in the equatorial plane due to the centrifugal potential.
According to Eq. (\ref{PROFILE}),
the density is constant on concentric ellipsoids
in the unperturbed rotating background.
Note that the normalization of the wave-function of the condensate
implies
\be
\int_V{d^3x\,\rho(\xvec)} = M,
\ee
where $M$ is the total mass in the cloud.

\subsection{Second harmonic modes of oscillation}

Next consider perturbations from the equilibrium state
of uniform rotation, with the spin vector along the $x_3$ axis.
Surface deformations related to various modes can be
classified by corresponding terms of the
expansion in surface harmonics labeled by indexes $l,m$.
Second order harmonic deformations correspond to $l=2$
with five distinct values of $m$, $-2\le m \le 2$.
The 18 equations represented by
Eq. (\ref{eq:v4}) separate into two independent
subsets which  are odd and even with respect to the index 3.
The corresponding oscillation modes can be treated separately.

\subsubsection{Transverse-shear modes}

These modes correspond
to the  surface deformations with $\vert m\vert = 1$
and represent relative shearing of the northern
and southern hemispheres of the spheroid.
The components of Eq. (\ref{eq:v4}), which are odd in
index 3, are:
\baray\label{mac_modes:1.1}
\lambda^2 V_{3;1} &=& - A_{3} V_{31} ,\\
   \label{mac_modes:1.2}
\lambda^2 V_{3;2} &=& - A_{3} V_{32} ,\\
   \label{mac_modes:1.3}
\lambda^2 V_{1;3} - 2 \Omega\lambda V_{2;3}
&=&-V_{13} +\Omega^2 V_{13} ,\\
   \label{mac_modes:1.4}
\lambda^2 V_{2;3} + 2 \Omega\lambda V_{1;3}
&=&-V_{23} + \Omega^2 V_{23}.
\earay
We sum the Eqs. (\ref{mac_modes:1.1}), (\ref{mac_modes:1.3})
and (\ref{mac_modes:1.2}), (\ref{mac_modes:1.4}), respectively, and
use the symmetry properties of $V_{ij}$ combined with Eqs.
(\ref{mac_modes:1.2}), (\ref{mac_modes:1.4}). This results in the
relations
\baray
&&\lambda\left(\lambda^2 + A_3 + 1 -\Omega^2\right) V_{13}
 - 2 \Omega (\lambda^2 + A_3) V_{23} = 0,
\label{mac_modes:1.5}\\
&&\lambda(\lambda^2+ A_3 + 1 - \Omega^2) V_{23}
-2\Omega  (\lambda^2 + A_3) V_{23} = 0.
\label{mac_modes:1.6}
\earay
The characteristic equation can be factorized by substituting
$\lambda = i\sigma$ and we find
\be\label{eq:rslt1}
\sigma^3 - 2\Omega \sigma^2 - (1 + A_3 - \Omega^2) \sigma
+ 2 \Omega A_3 =0  .
\ee
The roots are give by
\baray\label{sig1}
\sigma_1 = \frac{2\,\Omega}{3} + (s_++s_-) , \quad\quad
\sigma_{2,3} = \frac{2\,\Omega}{3} -\frac{1}{2} (s_+ +s_-)
    \pm \frac{i\sqrt{3}}{2} (s_+-s_-)  ,
\earay
where
\baray
\label{sig_coff}
s_{\pm}^3 &=&
\frac{\Omega}{3} \left(1 - 2 A_3 -\frac{\Omega^2}{9}\right)
\mp\frac{1}{9}\left[
\left(1+A_3+\frac{\Omega^2}{3}\right)^3 -\Omega^2
\left(1-2A_3-\frac{\Omega^2}{9}\right)^2\right]^{1/2}.
\earay
Three complementary modes follow from  Eqs. (\ref{sig1})-(\ref{sig_coff})
via the replacement  $\Omega\to -\Omega$.
It can be verified that the condition
\be
\left[\frac{\Omega}{3} \left(1-2A_3 -\frac{\Omega^2}{9}\right)\right]^2
 -\left[\frac{1}{3} \left(1+A_3 +\frac{\Omega^2}{3} \right)\right]^3 \le 0,
\ee
is satisfied for any $A_3$ and $\Omega < 1$, therefore all three
roots are real. The real frequencies of the transverse-shear modes
are shown in Fig. 1 as a function of $A_3$.  Note that
in the zero-temperature limit, to which present analysis is restricted,
these modes are purely real, i. e. represent undamped oscillations.

\subsubsection{Toroidal modes}
These modes correspond to $\vert m\vert = 2$ and
the motions in this case are confined to the planes parallel to the
equatorial plane. The components of Eq. (\ref{eq:v4}), which are
even in index 3, are:
\baray \label{mac_modes:2.1}
\lambda^2V_{3;3}&=& \delta\Pi -  A_3 V_{33}  ,\\
    \label{mac_modes:2.2}
\lambda^2V_{1;1}-2\Omega\lambda V_{2;1}
&=& \delta\Pi+(\Omega^2 - 1) V_{11} , \\
     \label{mac_modes:2.3}
\lambda^2V_{2;2}+2\Omega\lambda V_{1;2}
&=&  \delta\Pi+(\Omega^2 - 1) V_{22} ,\\
     \label{mac_modes:2.4}
\lambda^2V_{1;2}-2\Omega\lambda V_{2;2}
&=& (\Omega^2-1) V_{12} ,\\
     \label{mac_modes:2.5}
\lambda^2V_{2;1}+2\Omega\lambda V_{1;1}
&=& (\Omega^2-1) V_{21} .
\earay
We add Eqs. (\ref{mac_modes:2.4}) and   (\ref{mac_modes:2.5}),
and subtract Eqs. (\ref{mac_modes:2.2}) and   (\ref{mac_modes:2.3})
to find the following coupled equations:
\baray  \label{mac_modes:2.6}
 \left[\lambda^2 + 2 (1- \Omega^2)\right] (V_{11}-V_{22})
-4 \Omega \lambda V_{12} &=& 0 , \\\
     \label{mac_modes:2.7}
 \left[\lambda^2 + 2 (1- \Omega^2)\right] V_{12}
+\Omega\lambda (V_{11}-V_{22}) &=& 0  .
\earay
The characteristic equation for the toroidal modes is
\baray  \label{mac_modes:2.8}
&&\left[\lambda^2 - 2 (\Omega^2-1)\right]^2
+ 4 \lambda^2\Omega^2 = 0 ,
\earay
which is factorized by writing $\lambda=i\sigma$.
Two solutions are then
\be\label{eq:sig_tor}
\sigma_{1,2}=\Omega\pm\sqrt{2 - \Omega^2}.
\ee
There are two complementary modes which are found by substituting
$-\Omega$ for $\Omega$. Since the rotation frequency is bounded
($\Omega < 1$) the toroidal modes are stable independent of the
magnitude of the deformation in the equatorial plane.
For the same reason these modes are undamped in
the zero-temperature limit considered here.

\subsubsection{Pulsation modes}

To find the pulsation modes, which correspond to $m=0$,
we first add the  Eqs.
(\ref{mac_modes:2.2})-(\ref{mac_modes:2.3}), and then use
the Eq. (\ref{mac_modes:2.1}) to eliminate $\delta\Pi$ in the
result. In this manner we find that
\baray \label{mac_modes:3.1}
\left(\lambda^2/2-\Omega^2+ 1\right)(V_{11}+V_{22})
+ 2 \Omega\lambda (V_{1;2}-V_{2;1})
-(\lambda^2+ 2 A_3) V_{33}=0.
\earay
Subtracting Eqs. (\ref{mac_modes:2.5}) and (\ref{mac_modes:2.4})
(and discarding the $\lambda = 0$ root) one finds
\baray
    \label{mac_modes:3.2}
&&\lambda  (V_{1;2}-V_{2;1})
- \Omega (V_{11}+V_{22}) = 0 .
\earay
Eqs. (\ref{mac_modes:3.1}) and (\ref{mac_modes:3.2}) can be further
combined to a single equation:
\baray\label{mac_modes:3.3}
&&\left(\lambda^2+ 2 \Omega^2 +2\right) (V_{11}+V_{22})
-2(\lambda^2 +2 A_3) V_{33} =0.
\earay
It is instructive first to consider the case where the superfluid
is incompressible. Then the Eq. (\ref{mac_modes:3.3})
should be supplemented by the divergence free condition
\be
\frac{V_{11}+V_{22}}{a_1^2} + \frac{V_{33}}{a_3^2} = 0.
\label{DIV}
\ee
Again, writing $\lambda = i\sigma$, we find for the square
of frequency of the pulsation mode in the incompressible limit
\be
\sigma_0^2=\left(\frac{1}{2}+\frac{a_3^2}{a_1^2}\right)^{-1}
\, \left(1+\Omega^2 + 2 \frac{a_3^2}{a_1^2}A_3\right).
\label{incompuls}
\ee
As $\sigma_0^2$ is always positive, these modes
correspond to undamped stable oscillations.

For compressible fluids we need the variation of the
pressure tensor, which for adiabatic perturbations can be
written as
\be
\delta \Pi = (\gamma -1) \int\!d^3x~ \xi_i~\nabla_i p ,
\ee
where we assumed a polytropic equation of state
$p=K\rho^{\gamma}$; the polytropic index for a Bose gas
to leading order in the parameter $\rho a^3$ is equal 2.
To evaluate the gradient of the pressure we turn to
the Euler Eq. (\ref{eq:euler}) in the unperturbed limit,
\be
{\partial h\over\partial x_i}\equiv{1\over\rho}{\partial p\over
\partial x_i} = -\frac{1}{2}
{\partial\phi_{\rm tr}\over\partial x_i}
+\frac{1}{2}{\partial\vert\Omvec\crossprod\xvec\vert^2
\over\partial x_i},
\label{eq:euler:eq}
\ee
where the enthalpy $h$ is defined by $dh=dp/\rho$ and is
simply $h=\gamma K\rho^{\gamma-1}/(\gamma-1)$ for
$p=K\rho^\gamma$.
Substituting the explicit expression for the trapping
potential, we find that the gradient of the enthalpy
is a linear function of the coordinates and, hence,
the variation of the pressure tensor can be expressed in terms of
virials $V_{ii}$:
\be
\delta\Pi=-{(\gamma-1)\over 2}[(V_{11}+V_{22})(1-\Omega^2)+A_3V_{33}].
\label{delpi}
\ee
Any of the equations which are even in the index 3 now can
be used to close the system of equations for $V_{33}$ and the virial
combination $V_{11}+V_{22}$.
Substituting  Eq. (\ref{delpi}) for $\delta \Pi$ in,
e.g, Eq. (\ref{mac_modes:2.1}),
one finds
\be\label{mac_modes:3.4}
[\lambda^2+(\gamma+1)A_3]V_{33}+(\gamma-1)(1-\Omega^2)(V_{11}+V_{22})=0.
\ee
Eqs. (\ref{mac_modes:3.3}) and (\ref{mac_modes:3.4}) completely
determine the unknown virials;
the characteristic equation for the pulsation modes, which
is quadratic in $\lambda^2$, is
\baray
\lambda^4+\lambda^2[\gamma(A_3+2-2\Omega^2)+A_3+4\Omega^2]
+2A_3[\gamma(3-\Omega^2)+3\Omega^2-1]=0.
\earay
On substituting $\lambda = i\sigma$, the solution of the resulting
quadratic equation becomes
\baray \label{eq:PULS}
\sigma^2_\pm &=&\frac{1}{2}
\left[\gamma(A_3+2-2\Omega^2)+A_3+4\Omega^2\right]
\nonumber\\
&\pm&\frac{1}{2}
\sqrt{[\gamma(A_3+2-2\Omega^2)+A_3+4\Omega^2]^2-8A_3
[\gamma(3-\Omega^2)+3\Omega^2-1]}.
\earay
It is easy to see that there are only unstable modes
if $\gamma<(1-3\Omega^2)/(3-\Omega^2)\leq 1/3$; otherwise
all modes are stable. In particular, all modes are stable for
$\gamma=2$. Note, too, that there are twice as many solutions
as were found in the incompressible case; Eq. (\ref{incompuls})
can be recovered from Eq. (\ref{eq:PULS}) by taking the
$\gamma\to\infty$ limit of $\sigma^2_-$. The origin of the
additional root, $\sigma_+^2$, may be traced to the
$\lambda^2$ dependence of Eq. (\ref{mac_modes:3.4}), which
only reduces to the incompressibility condition, Eq. (\ref{DIV}),
in the $\gamma\to\infty$ limit if it is assumed that
$\vert\lambda^2\vert=\vert\sigma^2\vert\ll\gamma$.

The square of the pulsation modes as a function of the
$A_3$  is plotted in Fig. 3 for $\gamma = 2$, the polytropic index
of a dilute Bose gas.

\section{Equilibrium and oscillations in non-axisymmetric traps}

\subsection{Equilibrium shape}

When the symmetry with respect to the rotation axis is broken,
the equilibrium constraint Eq. (\ref{eq:figures}) needs to be
supplemented by a relation fixing the semi-axis ratios in the plane
orthogonal to the spin axis.
The diagonal components of Eq. (\ref{eq:eqbr}) provide the
triangle relations which determine the non-axisymmetric
equilibrium figure
\be
a_1^2(\Omega^2-\omega_0^2A_1) = a_2^2(\Omega^2-\omega_0^2A_2)
=-a_3\omega_0^2 A_3.
\ee
These simultaneous constraints can be written in a equivalent form
\be\label{eq:nonfigures}
\frac{a_3^2}{a_1^2} = \frac{1}{A_3}
\left(1-\frac{\Omega^2}{\omega_0^2}\right),\quad
\frac{a_3^2}{a_2^2} = \frac{1}{A_3}
\left(A_2-\frac{\Omega^2}{\omega_0^2}\right).
\ee
Note that in addition to the upper bound on the rotation frequency
set by the first relation (as in the case of the axisymmetric traps),
the second relations places a lower bound on the deformation
in the plane orthogonal to the spin axis: $A_2 \ge \Omega^2/\omega_0^2$.
Given the experimentally controlled values of $\Omega$, $A_3$ and $A_2$,
relations (\ref{eq:nonfigures}) determine, in a unique manner, the semi-axis
ratios of the resulting figure.

\subsection{Second harmonic modes of oscillation}

The non-axisymmetric modes can be found from Eq. (\ref{eq:v4}) in a manner
similar to the axisymmetric modes; however, now the degeneracy in indexes
1 and 2 should  be relaxed. The oscillation modes separate into two
non-combining groups which have even or odd parity with respect to the
index 3. Below, we shall treat these modes separately.

\subsubsection{Odd modes}
Among the four components  of Eq. (\ref{eq:v4}), which are odd in
index 3, three are identical to Eqs. (\ref{mac_modes:1.1}),
(\ref{mac_modes:1.2}), and (\ref{mac_modes:1.3}) under
non-axisymmetric conditions;  the component which is modifed reads
\baray
   \label{jac_modes:1.4}
\lambda^2 V_{2;3} + 2 \Omega\lambda V_{1;3}
&=&-A_2V_{23} + \Omega^2 V_{23}.
\earay
Summing Eqs. (\ref{mac_modes:1.1}), (\ref{mac_modes:1.3}),  and
(\ref{mac_modes:1.2}), (\ref{jac_modes:1.4}), we arrive at
\baray
&&\lambda\left(\lambda^2 + A_3 + 1 -\Omega^2\right) V_{13}
 - 2 \Omega (\lambda^2 + A_3) V_{23} = 0,
\label{jac_modes:1.5}\\
&&\lambda(\lambda^2+ A_3 + A_2  - \Omega^2) V_{23}
-2\Omega  (\lambda^2 + A_3) V_{23} = 0.
\label{jac_modes:1.6}
\earay
The sixth order characteristic equation derived from this algebraic system
is
\be
\lambda^6  + \left[1 + A_2 + 2(A_3 + \Omega^2)\right]\lambda^4 +
\left[A_2 + A_3 + A_2 A_3 + A_3^2 -(1
+A_2 - 6 A_3 )\Omega^2 + \Omega^4\right]\lambda^2
+ 4 A_3^2 \Omega^2=0.
\ee
The corresponding modes appear as conjugate pairs, i.e. there are only 3
distinct modes. The real parts of these modes (the modes are purely real)
are show in Fig. 3 as a function of $A_3$
for several values of $\Omega$ and fixed ratio $\Omega^2/A_2 = 0.1$.

\subsubsection{Even modes}

The even parity components of the Eq. (\ref{eq:v4}) which are
given by the Eqs. (\ref{mac_modes:2.1}),
(\ref{mac_modes:2.2}), and (\ref{mac_modes:2.4})
remain unchanged when axisymmetry is relaxed; the remainder
equations read
\baray
     \label{jac_modes:2.3}
\lambda^2V_{2;2}+2\Omega\lambda V_{1;2}
&=&  \delta\Pi+(\Omega^2 - A_2) V_{22} ,\\
     \label{jac_modes:2.4}
\lambda^2V_{2;1}+2\Omega\lambda V_{1;1}
&=& (\Omega^2-A_2) V_{21} .
\earay
In the incompressible limit these equations should be supplemented
by the condition (\ref{DIV}) (with a obvious modification of the second
term). In the compressible case the variations of the pressure become
\be
\delta\Pi=-{(\gamma-1)\over 2}[(1-\Omega^2)V_{11}
+(A_2-\Omega^2) V_{22}+A_3V_{33}].
\label{delpi2}
\ee
Using this relation in Eq. (\ref{mac_modes:2.1}) we find
\be\label{eq:delpi_bis}
\left[\lambda^2 + (\gamma+1) A_3\right] V_{33}
+(\gamma-1)\left[(1-\Omega^2)V_{11}+(A_2-\Omega^2)V_{22}\right] =0.
\ee
Equations (\ref{mac_modes:2.1}), (\ref{mac_modes:2.2}), (\ref{mac_modes:2.4})
and (\ref{jac_modes:2.3}),(\ref{jac_modes:2.4}) can be manipulated to the
the following set
\baray
\left[\lambda^2+2\left(1-\Omega^2\right)\right]V_{11}
-\left[\lambda^2+2\left(A_2-\Omega^2\right)\right]V_{22}-4\lambda\Omega V_{12}=0,
\\
\left[\lambda^2+2\left(1-\Omega^2\right)\right]V_{11}
+\left[\lambda^2+2\left(A_2-\Omega^2\right)\right]V_{22}-2(\lambda^2+2A_3)
V_{33}+ 4\lambda\Omega\left(V_{1;2}-V_{2;1}\right)=0,
\\
\left(\lambda^2+1 + A_2-2\Omega^2\right)V_{12} + \Omega\lambda (V_{11}-V_{22})=0,
\\
\lambda^2(V_{1;2}-V_{2;1}) -\Omega\lambda(V_{11}+V_{22})+(1-A_2)V_{12}=0,
\earay
which should be supplemented with Eq. (\ref{eq:delpi_bis}). The corresponding
characteristic equation is of order 8 and has been solved numerically. The
results for the real parts of the modes (which are purely real)
are shown in Fig. 4 as a function  of $A_3$ for several values
of $\Omega$ for andx fixed ratio $\Omega^2/A_2 = 0.1$.

\section{Conclusions}

We have analyized the hydrodynamic oscillations of Bose-condensed
atomic clouds at zero temperature in the Thomas-Fermi approximation.
The equilibrium shape of the cloud in a axisymmetric trap represent
either a prolate or oblate spheroid of revolution, which, for a particular
choice of the rotation rate and trap potential,
degenerates into a rotating sphere. The rotation frequency of the
condensate is bounded from above by the characteristic frequency of the harmonic oscillator
in a given trap ($\Omega^2/\omega_0^2 \le 1$).
We have also analyized  non-axisymmetric, triaxial ellipsoidal
figures, which admit equilibrium solutions under additional constrain
on the deformation in the plane orthogonal to the rotation axis $A_2 \ge
\Omega^2/\omega_0^2$.

Small amplitude oscillations have been derived for linear
perturbations from the rigidly rotating equilibrium background state.
The oscillations in axisymmetric traps, which are  related to the
transverse-shear and toroidal modes are found to be stable
for all values of the trap deformation and its rotation frequency.
These modes represent undamped oscillations
(all eigenfrequencies are real) in the absence of dissipation,
i.e., they are dynamically stable.
A dynamical instability against the quasi-radial
pulsation mode can arise when  $\gamma\leq(1-3\Omega^2)/(3-\Omega^2)\leq 1/3$. Otherwise
the system is stable, including the limit of the incompressible superfluid
($\gamma \to \infty)$, independent of the shape of equilibria.
For a Bose gas with $\gamma = 2$ the above stability condition is satisfied;
hence, these modes represent stable oscillations independent of
the rotation rate and deformation.
Numerical results for the oscillations in non-axisymmetrical
traps show that, both,  the even and odd partity second harmonic modes
of oscillations of triaxial clouds  are stable for the adiabatic index $\gamma = 2$
relevant to the interacting dilute Bose gas.

Our results for axisymmetric traps (in which case analytical results are available) reduce
to those of  Ref. \cite{STRINGARI} when the rotation frequency is
set to zero and $\gamma =2$. Indeed, from Eqs. (\ref{eq:rslt1}), (\ref{eq:sig_tor}) and
(\ref{eq:PULS}), which describe to the oscillations with $m = \pm 1$,
$m = \pm 2$ and $m = 0$, respectively, for $l = 2$, we find
\baray
\sigma(l = 2, m =\pm 1) &=& \pm \sqrt{1+A_3},\\
\sigma(l = 2, m =\pm 2) &=& \pm \sqrt{2,}\\
\sigma(l = 2, m = 0) &=& 2 + \frac{3A_3}{2} \pm\frac{1}{2} \sqrt{ 16 - 16 A_3 + 9 A_3^2},
\earay
which coincide with Eqs. (22), (23) and (24) of Ref. \cite{STRINGARI}
on making the appropriate changes in the parameters describing
the trapping potential.

It is instructive also to compare the results above
to the classical analysis of the equilibrium and stability of the
self-gravitating fluids\cite{CHANDRA} in view of speculations\cite{ODELL}
that intense off-resonant laser beams can give rise
to a gravitational-type potential between the condensate
particles leading to self-bound configurations.
Self-gravitating systems, in particular the
axisymmetric figures, are stable against transverse-shear
and pulsation modes in the incompressible limit; the same is found
in the above model of trapped rotating condensates. However, in
the compressible case, the self-gravitating fluids are unstable
against the pulsation modes whenever the adiabatic index
$\gamma \le 4/3$ (the precise value of the critical $\gamma$
depends on the rotation rate) and are stable otherwise.
In the present case, the system is unstable when
$\gamma < (1-3\Omega^2)/(3-\Omega^2)$, where the rotation
frequency covers the range $\Omega_{c1}\le \Omega \le 1$.
Another major difference between the
two systems is their stability against the toroidal modes.
The axisymmetric self-gravitating fluids are unstable dynamically
(i.e. in the absence of dissipation) beyond the
point $a_3^2/a_1^2 = 0.05$,  as the deformation is increased.
At a smaller deformation $a_3^2/a_1^2 = 0.19$, these oscillations become
neutral, which is a prerequisite of the onset of
secular (i.e. driven by the viscosity) instability.
In contrast, the toroidal modes of the trapped
condensates are always dynamically stable and there are no neutral points within the allowed parameter
space where equilibrium figures exist.

The present model can be extended for a  study of the
higher order ($l > 2$) harmonic oscillations by constructing
higher order virial equations as well as to
finite temperatures, in which case the  viscosity of
thermal excitation, hence secular instabilities,
and  mutual friction between the vortex lattice state and the excitations
play a role\footnote{Such a program for self-gravitating superfluids
has been carried out recently in  Ref. \cite{SW}.}.

\acknowledgements
We acknowledge the generous hospitality and support of the
European Center for Theoretical Studies in Nuclear Physics
and Related Areas (ECT$^*$), Trento, were this work was
carried out. We are grateful to Professor B. Mottelson for a
helpful conversation.
A.S. acknowledges the support of the
Nederlandse Organisatie voor Wetenschappelijk Onderzoek
via  the Stichting voor Fundamenteel Onderzoek der Materie.
I.W. acknowledges partial support from  NASA.

\appendix
\section*{Hydrodynamic equations of motion for rotating condensate}

As is well-known, dilute Bose gases can be described
in the mean-field approximation in terms of the Gross-Pitaevskii
(G-P) theory. The latter theory (in analogy to
the phenomenological theories of superconductivity and of
superfluid He$^4$ near the $\lambda$-point), is based on a
Ginzburg-Landau type functional for the wave-function $\psi$ of the
coherent state, whose variation provides the equation of motion
for $\psi$. In the stationary case this functional for a dilute Bose gas
has the well-known form
\be
E(\psi) =\int\left[\frac{\hbar^2}{2m}\vert\bnabla\psi(\xvec)\vert^2
+\frac{m}{2}\phi_{\rm tr}(\xvec)\vert\psi(\xvec)\vert^2+\frac{1}{2} U_0 \vert\psi(\xvec)\vert^4
\right] d^3x  ,
\ee
where $U_0\equiv 4\pi\hbar^2a/m$. The G-P equation is obtained by taking
functional derivative with respect to $\psi^*$
subject to the constarint that the particle number $N$ is constant.
The extremum condition $\delta (E-\mu N)/\delta\psi^* =0$ gives
\be\label{GP}
-\frac{\hbar^2}{2m}\bDelta\psi(\xvec)
+\frac{1}{2}\left[m\phi_{\rm tr}(\xvec)+ U_0\vert\psi(\xvec)\vert^2
\right] \psi(\xvec) = \mu\psi(\xvec),
\ee
where the Lagrangian multiplier $\mu$ has the meaning of
the chemical potential of particles. The
time-dependent generalization of (\ref{GP}) follows on the assumption
that the temporal variations  of $\psi$ should be described by a first
order equation which, by analogy with the quantum mechanics, is written as
\be\label{TDGP}
i\hbar \frac{\partial \psi(\xvec ,t)}{\partial t} =
-\frac{\hbar^2}{2m}\bDelta\psi(\xvec ,t)
+\left[\frac{m}{2}\phi_{\rm tr}(\xvec)+ U_0\vert\psi(\xvec ,t)\vert^2\right] \psi(\xvec ,t).
\ee
On writing
$
\psi(\xvec ,t) = \eta(\xvec ,t) e^{i\varphi(\xvec ,t)},
$
the superfluid density  and velocity can be expressed as
\be \label{SUBST}
\rho(\xvec ,t) = m\vert \psi(\xvec ,t)\vert^2 = m\eta(\xvec ,t)^2, \quad
\vvec(\xvec ,t) = \frac{\hbar}{m} \bnabla \varphi(\xvec ,t).
\ee
The real part of Eq. (\ref{TDGP}) after deviding by $m\eta$, applying
a gradient, and multiplying the result by $\rho$ becomes
\be\label{eq:preeuler}
\rho \left(\frac{\partial}{\partial t} + \vvec\bnabla\right) \vvec
+\rho \bnabla\left(-\frac{\hbar^2}{2m^2}\frac{\Delta\eta}{\eta}
+\frac{1}{2} \phi_{\rm tr}+\frac{1}{m} U_0\eta^2\right) = 0,
\ee
which is the Euler equation for the condensate in the zero-temperature,
inviscid limit. It differs from the analogous equation for the ordinary
fluids by the ``quantum pressure'' term $\propto \Delta \eta/\eta$. Note that
any constant term can be added to the second bracket, for example, the
chemical potential in the ground state.

The imaginary part of Eq. (\ref{TDGP}), on multiplying by
$2 m \eta/\hbar$, leads to the mass conservation equation
\be\label{eq:euler_rho}
\frac{\partial}{\partial t}\rho + \bnabla \rho \vvec = 0.
\ee
When gradients of the $\psi$ are small,  the ``quantum pressure'' term
in  Eq. (\ref{eq:preeuler}) can be dropped and it
reduces to
\be\label{eq:euler2}
\rho \left(\frac{\partial}{\partial t} + \vvec\bnabla\right) \vvec=
- \bnabla p -\frac{\rho}{2} \bnabla \phi_{\rm tr}
\ee
where $ p \equiv (U_0/2m)\rho^2 = (2\pi\hbar^2 a/m^2) \rho^2$. Note that
{\it formally} Eq. (\ref{eq:euler2}) is identical to the Euler equation
in the ordinary hydrodynamics; the distinctive feature of the superfluid is
that the superflow is irrotational $\bnabla\times \vvec =0$ in general
(the special case of the rotating superfluid, when the analogy becomes
perfect, is discussed below).

Eq. (\ref{eq:euler2}) can be derived also starting from the momentum conservation
equation:
\be\label{eq:momentum}
\frac{i\hbar}{2}\frac{\partial}{\partial t}
\left[\psi(\xvec ,t) \bnabla \psi^*(\xvec ,t) -
\psi^*(\xvec ,t) \bnabla \psi(\xvec ,t)
 \right] + \frac{\partial}{\partial x_k} \Pi_{ik}= \rho \frac{1}{2}
\bnabla\phi_{\rm tr} ,
\ee
where
\be
\Pi_{ik} = \frac{\hbar^2}{4m^2}
\left[\frac{\partial\psi}{\partial x_i}\frac{\partial\psi^*}{\partial x_k}
-\psi\frac{\partial^2\psi^*}{\partial x_i\partial x_k}+{\rm c.c.}
\right]+p\delta_{ik},
\ee
$p$ is the pressure,  c.c. stands for complex conjugate. The right-hand-side
of Eq. (\ref{eq:momentum}) is the external force per unit volume. On writing
$
\psi(\xvec ,t) = \eta(\xvec ,t) e^{i\varphi(\xvec ,t)},
$
and using the relations  (\ref{SUBST}) Eq.
(\ref{eq:momentum}) becomes
\be
\frac{\partial}{\partial t} \rho v_i +
\frac{\partial}{\partial x_k}\left(
\rho v_i v_k + p\delta_{ik}\right)= \rho \frac{1}{2}
\nabla_i\phi_{\rm tr} .
\ee
The time derivative of $\rho$ can be eliminated
in terms of Eq.  (\ref{eq:euler_rho}).
If we use for the pressure of the dilute Bose gas
the relation
\be
p = \frac{2\pi\hbar^2a\rho^2}{m^2}\left[1+\frac{64}{5}
\left(\frac{\rho a^3}{\pi m}\right)^{1/2} \right]
\ee
and keep the leading order term in the diluteness parameter,
we arrive again at Eq. (\ref{eq:euler2}).

If the condensate is rotating at a angular velocity which is larger
than the critical one $\Omega_{c1}$, its energy is minimized via creation of
vortices; then  the curl of the last relation in (\ref{SUBST}) is non-zero, rather
the  phase of the superfluid order parameter changes by $2\pi$ around a
path which encircles vortex lines
\be\label{curlv}
\bnabla\times  {\vvec} =\frac{\hbar}{m}
\bnabla\times   \bnabla \varphi(\xvec) = \frac{2\pi \hbar}{m}
\nuvec \sum_{j} \delta^{(2)} ({\xvec}-{\xvec}_{j}),
\ee
where $\nuvec$
is a unit vector along a vortex line, ${\xvec}_{j}$ is the
radius vector of a vortex line in the plane  orthogonal to the vector
$\nuvec$, and $\delta^{(2)}$ is a two-dimensional delta
function in this plane. If $\Omega \gg \Omega_{c1}$ the macroscopic
hydrodynamic equations involve only {\it course-grained} quantities,
which are averages over a large number of vortices (i.e. over
scales much larger than the size of a single vortex). The
right-hand side of Eq. (\ref{curlv}) then becomes proportional to the
density of the vortex lines, $n_v$, since the continuum limit
of vortex distribution implies
$\sum_{j} \delta^{(2)} ({\xvec}-{\xvec}_{j}) = n_v $. The left-hand side of
Eq. (\ref{curlv}) in the course-grained limit gives $2\Omega$, since  the
energy is minimized by a superflow which mimics a rigid body rotation
(this minimization is carried out e.g. , in e.g. Ref.
\cite{KHALATNIKOV}).
Writing Eq. (\ref{eq:euler2}) in the
frame rotating uniformly with the angular velocity $\Omvec$ amounts
to adding to the right-hand side of this equation the centrifugal potential
$\vert\Omvec\times \xvec\vert^2/2$ and the Coriolis acceleration
$2 \uvec\times \Omvec$ (here $\uvec$ is the superfluid velocity in the
rotating frame). With this substitution we recover Eq. (\ref{eq:euler}).
Note that the analogy to the Euler  equation for a uniformly
rotating ordinary fluid now is complete.

\newpage

\begin{center}
\includegraphics[height=5.2in,width=5.2in,angle=-90]{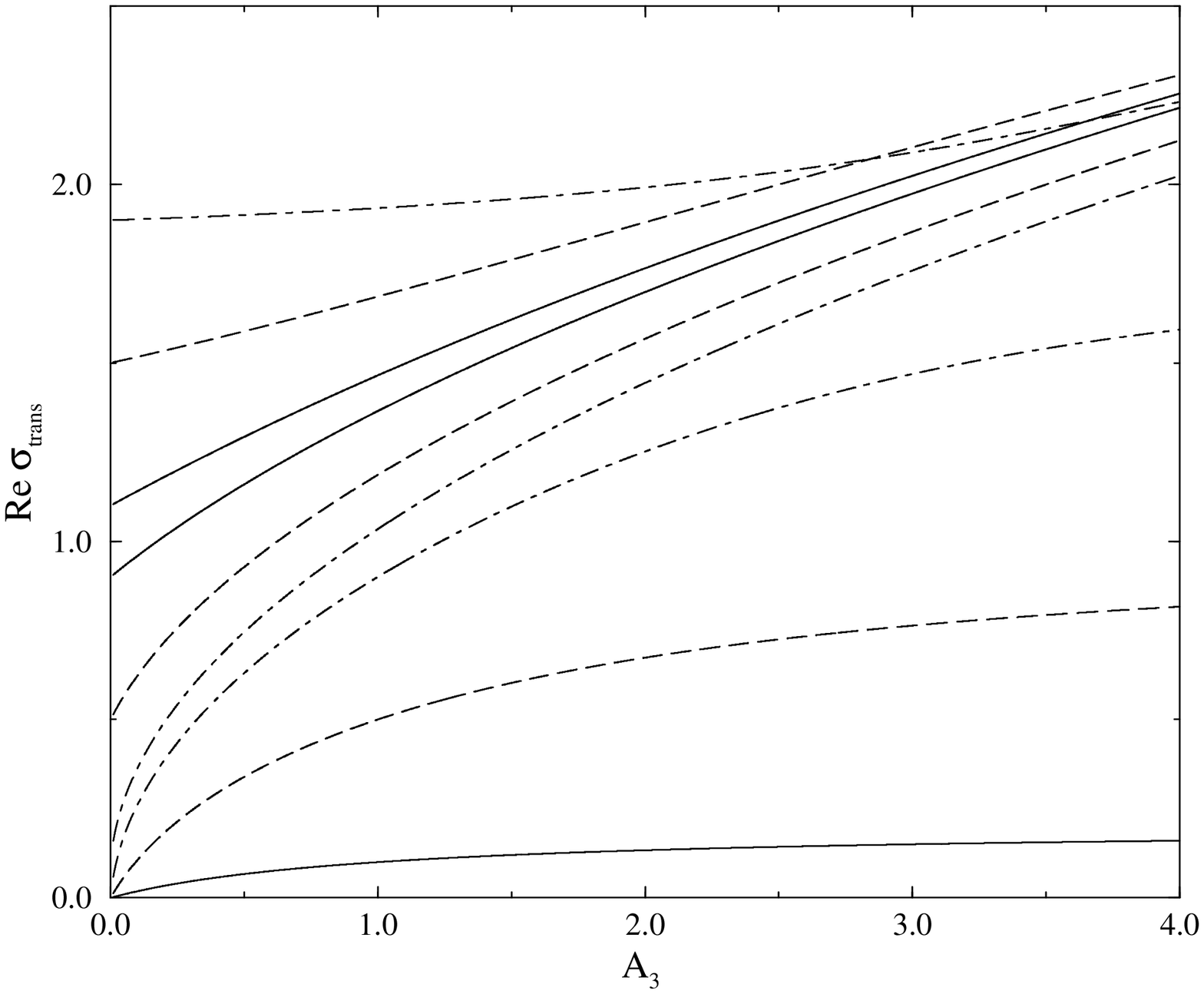}
\end{center}
{\footnotesize{Fig 1: The three real frequencies of the
transverse-shear modes in axisymmetric traps as a function of
deformation parameter $A_3$ for three values of $\Omega = 0.1$
({\it solid line}), 0.5 ({\it dashed line}), 0.9
({\it dashed-dotted line}); here the spin frequency is
measured in units of $\omega_0$.
}}
\label{fig1}

\begin{center}
\includegraphics[height=5.2in,width=5.2in,angle=-90]{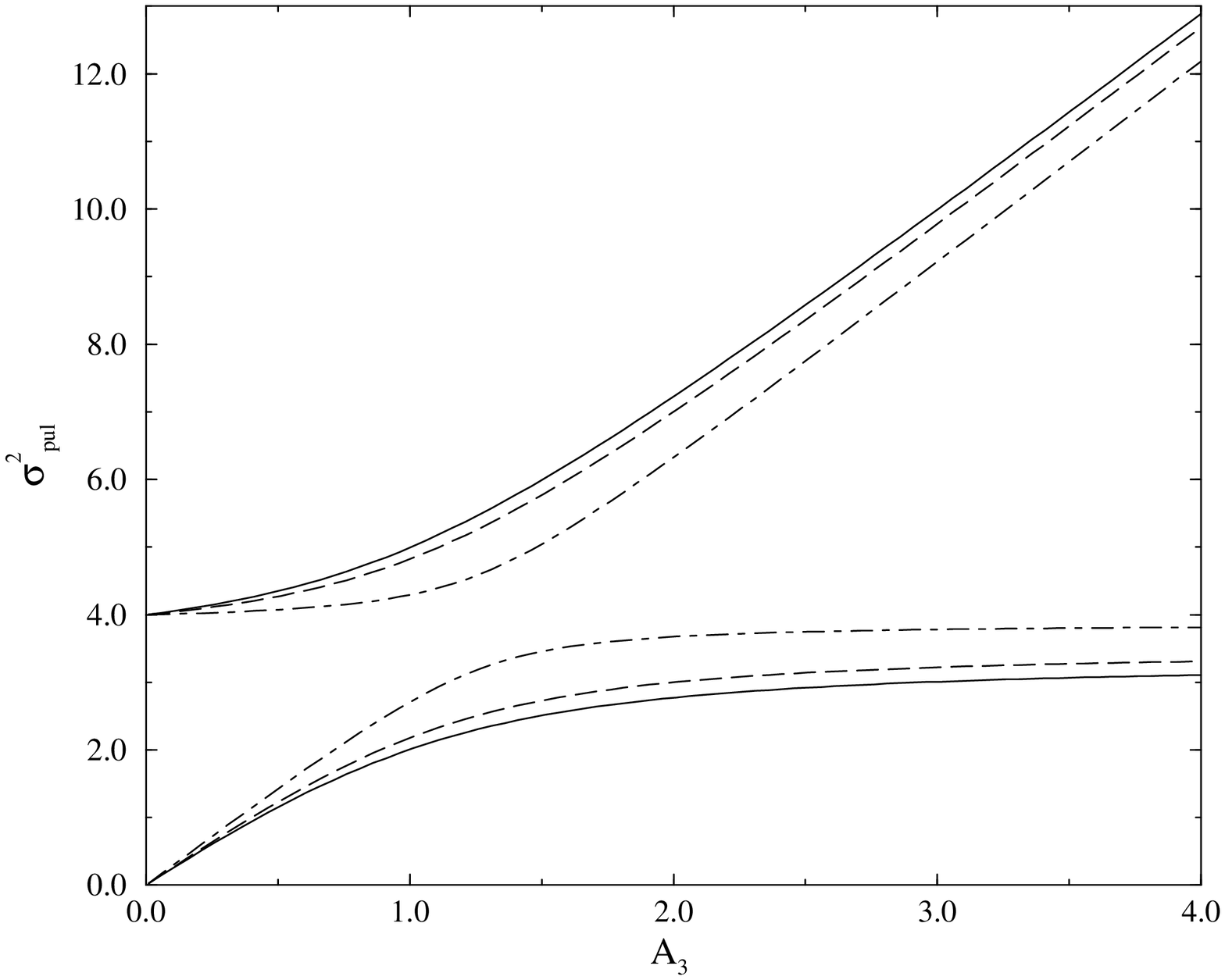}
\end{center}
{\footnotesize{Fig 2: The square of the pulsation
modes in axisymmetric traps for $\gamma = 2$.
Conventions are the same as in Fig. 1.
}}
\label{fig2}

\begin{center}
\includegraphics[height=5.2in,width=5.2in,angle=-90]{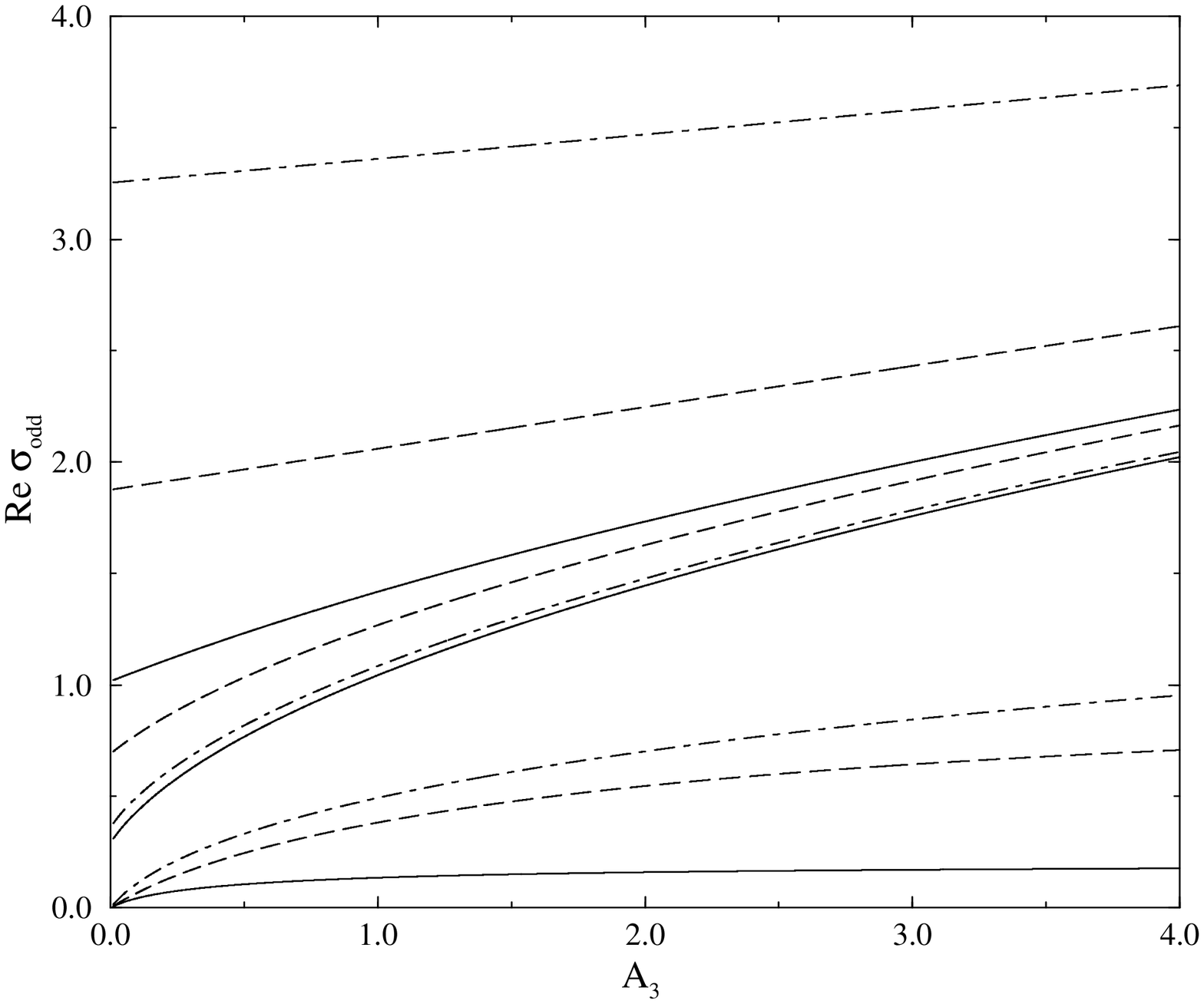}
\end{center}
{\footnotesize{Fig 3: The real parts of the odd parity modes
in non-axisymmetric traps for the fixed ratio $A_2/\Omega^2 = 0.1$.
Conventions are the same as in Fig. 1.
}}
\label{fig3}

\begin{center}
\includegraphics[height=5.2in,width=5.2in,angle=-90]{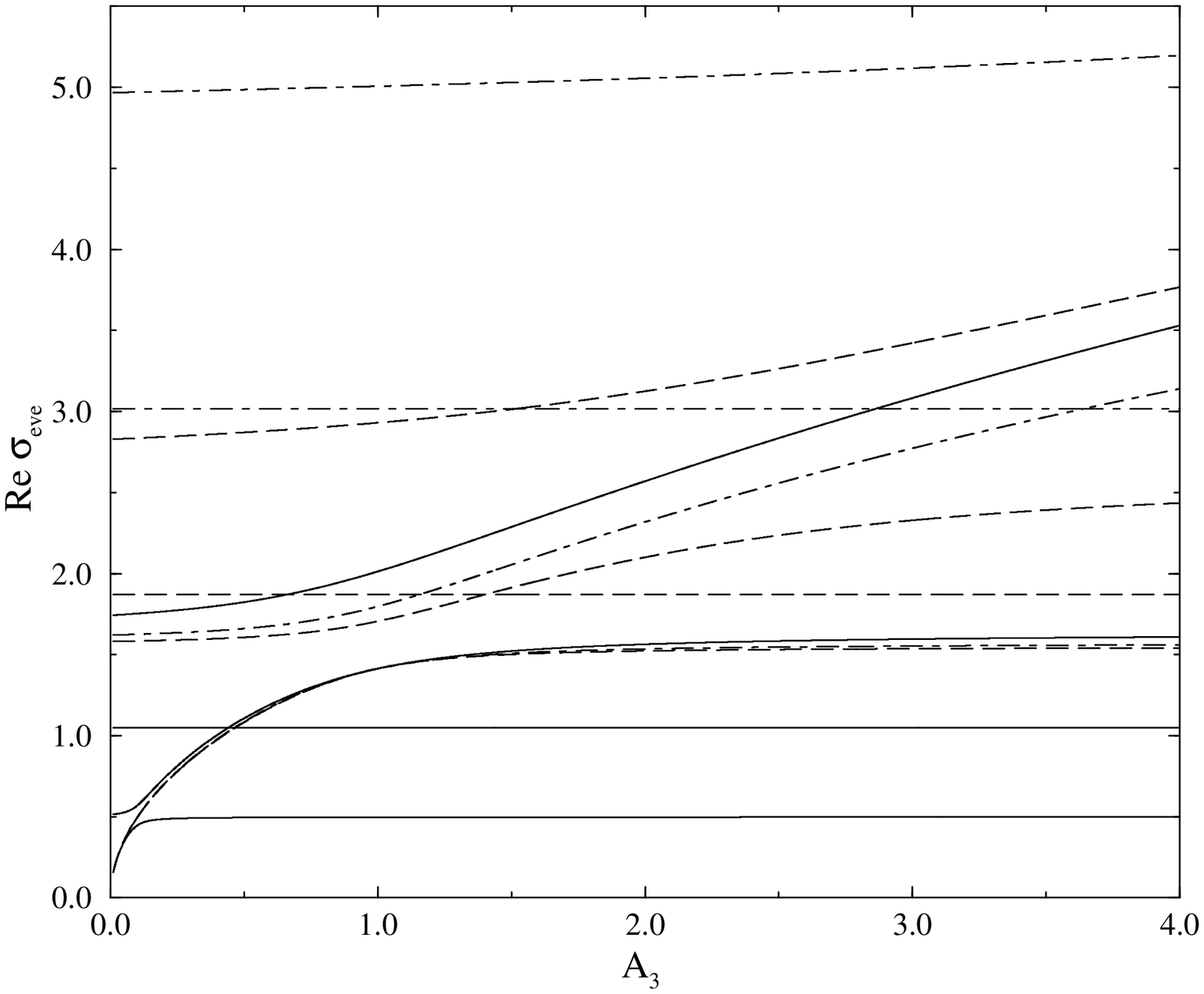}
\end{center}
{\footnotesize{Fig 4: The real parts of the even parity modes
in non-axisymmetric traps for the fixed ratio $A_2/\Omega^2 = 0.1$
and $\gamma = 2$.
Conventions are the same as in Fig. 1.
}}
\label{fig4}

\begin{references}
\bibitem{REMARK} It is worthwhile to note that Kerr's solution to
                 the equations of the general theory of relativity,
                 which describes a rotating black hole,  is
                 the only other known example of an equilibrium figure
                 which is a rotating sphere. In the present case, however, the
                 equilibrium spherical figure is maintained due to
                 the exact cancellation of the effect of the
                 centrifugal potential by the external trapping potential.
\end{references}
\end{document}